\documentstyle[12pt,psfig]{article}

\def\baselinestretch{1.2}
\begin{document}

\begin{titlepage}

\begin{flushright}
OHSTPY-HEP-T-00-003\\
hep-th/0003249
\end{flushright}
\vfil\vfil

\begin{center}

{\Large {\bf Towards a SDLCQ test of the Maldacena Conjecture}}

\vfil
J.R. Hiller,$^a$ O. Lunin,$^b$ S. Pinsky,$^b$ and U. Trittmann$^b$
\\

\vfil
$^a$Department of Physics\\ University of Minnesota Duluth \\
Duluth, MN~~55812~~USA\\
\vfil

$^b$Department of Physics \\ Ohio State University, Columbus, OH~~43210~~USA\\
\vfil
\end{center}

\begin{abstract}
\noindent We consider the Maldacena conjecture applied to the near horizon
geometry of a D1-brane in the supergravity approximation and present
numerical results of a test of the conjecture against the boundary field
theory calculation using DLCQ.  We previously calculated the two-point function
of the stress-energy tensor on the supergravity side; the methods of
Gubser, Klebanov, Polyakov, and Witten were used. On the field theory side, we
derived an explicit expression for the two-point function in terms of data
that may be extracted from the supersymmetric discrete light cone quantization
(SDLCQ) calculation at a given harmonic resolution. This yielded a well
defined numerical algorithm for computing the two-point function. For the
supersymmetric Yang-Mills theory with 16 supercharges that arises in the
Maldacena conjecture, the algorithm is perfectly well defined; 
however, the size
of the numerical computation prevented us from obtaining a numerical check of
the conjecture. We now present numerical results with approximately
1000 times as many states as we previously considered.  These results
support the Maldacena conjecture and are within $10-15\%$ of the
predicted numerical results in some regions. Our results are still
not sufficient to demonstrate convergence, and, therefore, cannot be considered
to a numerical proof of the conjecture. We present a method for using a
``flavor'' symmetry to greatly reduce the size of the basis and  discuss a
numerical method that we use which is particularly well suited for this type
of matrix element calculation.
\end{abstract}

\vfil\vfil\vfil
\begin{flushleft}
March 2000
\end{flushleft}

\end{titlepage}
\renewcommand{\baselinestretch}{1.05}  

\section{Introduction}

The discovery that certain field theories admit concrete realizations as
a string theory on a particular background has caused a great deal of
excitement in recent years
\cite{adscft}.  However, attempts to apply these correspondences to
study the details of these theories have only met with limited success
so far. The problem stems from the fact that our understanding of both
sides of the correspondence is limited. On the field theory side, most
of what we know comes from perturbation theory where we assume that the
coupling is weak. On the string theory side, most of what we know comes
from the supergravity approximation where the curvature is small.  There
are no known situations where both approximations are simultaneously
valid. At the present time, comparisons between the dual gauge/string
theories have been restricted to either qualitative issues or quantities
constrained by symmetry. Any improvement in our understanding of field
theories beyond perturbation theory or string theories beyond the
supergravity approximation is, therefore, a welcome development.

Previously \cite{ahlp99} we showed that Supersymmetric Discrete Light
Cone Quantization (SDLCQ) of field theories
\cite{dlcqpapers,BPP,lup99,mss95} can, in principle, be used to make a
quantitative comparison with the supergravity approximation on the
string theory side of the correspondence.  We discussed this in two
space-time dimensions where the SDLCQ approach works particularly well;
however, it can in principle be extended to more dimensions.

We will study the field theory/string theory correspondence motivated by
considering the near-horizon decoupling limit of a D1-brane in type IIB
string theory \cite{imsy96}. The gauge theory corresponding to this
theory is the Yang-Mills theory in two dimensions with 16 supercharges.
Its SDLCQ formulation was recently reported in
\cite{alpp98}, and recent work  has put the use of SDLCQ for this class of
problems on a stronger footing \cite{bil99}. This is probably the
simplest known example of a field theory/string theory correspondence
involving a field theory in two dimensions with a concrete Lagrangian
formulation.

A convenient quantity that can be computed on both sides of the
correspondence is the correlation function of gauge invariant operators
\cite{GKP,Wit}. We will focus on two-point functions of the
stress-energy tensor.  This turns out to be a very convenient quantity
to compute for reasons that are discussed in \cite{ahlp99}.  Some
aspects of this, as it pertains to a consideration of black hole entropy,
were recently discussed in \cite{akisunny}. In the DLCQ literature, the
spectrum of hadrons is often reported \cite{lup99}.  This would be fine
for theories in a confining phase. However, we expect the SYM in two
dimensions to flow to a non-trivial conformal fixed point in the
infra-red.  The spectrum of states will therefore form a continuum and
will be cumbersome to handle.  On the string theory side, entropy
density and the quark anti-quark potential are frequently reported. The
definition of entropy density requires that we place the field theory in
a space-like box which is incommensurate with the light-like box of
DLCQ.  Similarly, a static quark anti-quark configuration does not fit
very well inside a discretized light-cone geometry.  A correlation
function of point-like operators does not suffer from these problems.

\section{Correlation functions in supergravity}

The correlation function of the stress-energy tensor on the string theory
side, with use of the supergravity approximation, was presented
in \cite{ahlp99},
and we will only quote the result here.  The computation is essentially a
generalization of \cite{GKP,Wit}.  The main conclusion on the
supergravity side was reported recently in \cite{akisunny}. Up to a
numerical coefficient of order one, which we have suppressed, we found
that
\begin{equation}
\langle {\cal O}(x) {\cal O} (0) \rangle = {N_c^{{3 \over 2}} \over
g_{YM} x^5}\,. \label{SG}
\end{equation}
This result passes the following important consistency test.  The SYM
in 2 dimensions with 16 supercharges have conformal fixed points in
both UV and IR with central charges of order $N_c^2$ and $N_c$,
respectively. Therefore, we expect the two point function of the
stress-energy tensor to scale like $N_c^2/x^4$ and $N_c/x^4$ in the deep UV and
IR, respectively. According to the analysis of \cite{imsy96}, we expect
to deviate from these conformal behaviors and cross over to a regime
where the supergravity calculation can be trusted. The crossover occurs
at $x = 1 / g_{YM} \sqrt{N_c}$ and $x = \sqrt{N_c} / g_{YM}$. At these
points, the $N_c$ scaling of (\ref{SG}) and the conformal result match
in the sense of the correspondence principle \cite{hop96}.

\section{Correlation functions in SUSY with 16 Super Charges}

The challenge then is to attempt to reproduce the scaling relation
(\ref{SG}), fix the numerical coefficient, and determine the details
of the crossover behavior using SDLCQ. In order to actually
evaluate the correlation functions, we must resort to numerical
analysis.

The technique of SDLCQ is reviewed in \cite{BPP},
so we will be brief here.  The basic idea of light-cone quantization
is to parameterize space-time using light-cone coordinates $x^+$ and
$x^-$ and to quantize the theory making $x^+$ play the role of time.
In the discrete light cone approach, we require the momentum $p_- =
p^+$ along the $x^-$ direction to take on discrete values in units of
$p^+/K$ where $p^+$ is the conserved total momentum of the system and
$K$ is an integer commonly referred to as the harmonic
resolution \cite{dlcqpapers}.
One can think of this discretization as a consequence of compactifying
the $x^-$ coordinate on a circle with a period $2L = {2 \pi K /
p^+}$. The advantage of discretizing on the light cone is the fact that
the dimension of the Hilbert space becomes finite.  Therefore, the
Hamiltonian is a finite dimensional matrix, and its dynamics can be
solved explicitly.  In SDLCQ one makes the DLCQ approximation to the
supercharges, and these discrete representations satisfy the
supersymmetry algebra. Therefore SDLCQ enjoys the improved
renormalization properties of supersymmetric theories.  Of course, to
recover the continuum result, we must send $K$ to infinity and as luck
would have it, we find that SDLCQ usually converges faster than the
naive DLCQ\@. Of course, in the process the size of the matrices will
grow, making the computation harder and harder.

Let us now return to the problem at hand. We would like to compute a
general expression of the form 
$F(x^-,x^+) = \langle {\cal O}(x^-,x^+) {\cal O} (0,0) \rangle$\@.
In DLCQ, where we fix the total momentum in the $x^-$ direction, it is
more natural to compute the Fourier transform and express the transform
in a spectral decomposed form
\begin{equation}
\tilde{F}(P_-,x^+) = {1 \over 2 L} \langle {\cal O}(P_-,x^+) {\cal O}(-P_-,
0) \rangle\ =\sum_i {1 \over 2 L} \langle 0| {\cal O}(P_-) | i
\rangle e^{-i P_+^i x^+} \langle i|  {\cal O}(-P_-,0) |0 \rangle\ .
\label{master}
\end{equation}
The position-space form of the correlation function is recovered by Fourier
transforming with respect to $P_- =K\pi/L$.  We can continue to Euclidean
space by taking $r = \sqrt{2 x^+ x^-}$ to be real. The result for the
correlator of the  stress-energy tensor was presented in
\cite{ahlp99}, and we only quote the results here:
\begin{equation}
F(x^-,x^+)=
\sum_i\left|{L \over \pi} \langle 0 | T^{++}(K) |i \rangle \right|^2
\left({x^+ \over x^-}\right)^2 {M_i^4 \over 8 \pi^2 K^3}
K_4\left(M_i\sqrt{2 x^+ x^-}\right)\,,
\end{equation}
where $M_i$ is a mass eigenvalue and $K_4(x)$ is the
modified Bessel function of order 4. In \cite{alpp98} we
found that the momentum operator $T^{++}(x)$ is given by
\begin{equation}
T^{++}(x) =  {\rm tr} \left[ (\partial_- X^I)^2 + {1 \over 2} \left(i
u^\alpha \partial_- u^\alpha  - i  (\partial_- u^\alpha) u^\alpha
\right)\right], \qquad I, \alpha = 1 \ldots 8\,
\label{susycorr}
\end{equation}
where $X$ and $u$ are the physical adjoint scalars and fermions
respectively, following the notation of \cite{alpp98}.  When
discretized, these operators have the mode expansions
\begin{eqnarray}
X_{i,j}^I & = & {1 \over \sqrt{4 \pi}} \sum_{n=1}^{\infty} {1 \over
\sqrt{n}} \left[ a^I_{ij} (n) e^{-i \pi n x^-/L} +
a^{\dagger I}_{ji}(n) e^{i \pi n
x^-/L} \right]\,,\nonumber \\
u_{i,j}^\alpha & = & {1 \over \sqrt{4 L}} \sum_{n=1}^{\infty} \left[
b^\alpha_{ij} (n) e^{-i \pi n x^-/L} + b^{\dagger\alpha}_{ji}(-n) 
e^{i \pi n x^-/L}
\right]\,.
\end{eqnarray}
The matrix element $(L/\pi) \langle 0 | T^{++}(K) | i \rangle$ is
independent
of $L$ and can be substituted directly  to give an
explicit expression for the two-point function. We see immediately that
the correlator has the correct small-$r$ behavior, for in that limit, it
asymptotes to
\begin{equation}
\left({x^- \over x^+ }\right)^2 F(x^-,x^+) =
   {N_c^2(2 n_b + n_f) \over 4  \pi^2 r^4}
            \left(1 - {1 \over K}\right)\,.
\end{equation}

On the other hand, the contribution to the correlator from strictly massless
states is given by
\begin{equation}
\left( x^- \over x^+ \right)^2 F(x^-,x^+) = 
             \sum_i\left| {L \over \pi} 
            \langle 0 | T^{++}(K) | i\rangle \right|^2_{M_i=0} 
                {6 \over K^3 \pi^2 r^4}\,.
\end{equation}
It is important that this $1/r^4$ behavior at large $r$ not be confused with
the $1/r^4$ behavior that we seek at large $r$. First of all, there is
not supposed to be any massless physical bound state in this theory, and,
secondly, it has the wrong $N_c$ dependence.

Relative to the $1/r^4$ behavior at small $r$, the $1/r^4$ behavior at
large $r$ that we expect is down by a factor of $1/N_c$. Since we are
doing a large-$N_c$ calculation, this behavior is suppressed. We can
only hope to see the transition from the $1/r^4$ behavior at small $r$ to the
region where the correlator behaves like $1/r^5$.

\section{Discrete Symmetries of the Problem.}

In order to calculate the correlation function we use the expression
(\ref{master}). This means that after diagonalizing the Hamiltonian $P^-$ one
should evaluate the projection of each eigenfunction on the specific state
$T^{++}(-K)|0\rangle$. The fact that we are only interested in states which
have nonzero value of such projection leads to significant simplifications.

One can diagonalize any of the eight supercharges $Q^-_{\alpha}$. In the
continuum limit, the result does not depend on the value of $\alpha$ that one
chooses, but in DLCQ the situation is a little more subtle. As was shown in
\cite{alpp98}, while the spectrum of $(Q_\alpha^-)^2$ is the same for all
$\alpha$, the
wave functions depend on the choice of supercharge. This dependence is an
artifact of DLCQ and should disappear in the continuum limit. We refer to
\cite{alpp98} for the discussion of this issue.  Here we will just pick one
supercharge (for example, $Q^-_1$). Since the state $T^{++}(-K)|0\rangle$
  is a singlet under R--symmetry acting on the ``flavor'' index of
$Q^-_{\alpha}$, the correlator (\ref{master}) does not depend on the choice of
$\alpha$ even at finite resolution.

A significant simplification occurs at this stage. Suppose there exists an
operator $S$ commuting with both $P^-$ and $T^{++}(-K)$ and such that
$S|0 \rangle=s_0|0 \rangle$. Then the Hamiltonian and $S$ can be diagonalized
simultaneously.  From now on we assume that the set of states
$|i \rangle$ is a result of
such diagonalization. In this case, only states satisfying the condition
$S|i \rangle=s_0|i \rangle$ contribute to the sum in (\ref{master}), and we
only need to diagonalize $P^-$ in this sector. So if one finds a large enough
set of appropriate operators $S$, then the size of the problem can be
significantly reduced. By looking at the structure of the state
$T^{++}(-K)|0\rangle$ one
can conclude, given arbitrary permutations $P$ and $Q$
of the $8$ flavor indices, that any transformation of the form
\begin{eqnarray}
a^I_{ij}(k)\rightarrow f(I)a^{P[I]}_{ij}(k), \qquad f(I)=\pm 1
 \nonumber  \\ 
b^\alpha_{ij}(k)\rightarrow g(\alpha)b^{Q[\alpha]}_{ij}(k), \qquad
g(\alpha)=\pm 1 
\label{initsymm}
\end{eqnarray}
commutes with $T^{++}(-K)$, and that the vacuum is an eigenstate of
this transformation with eigenvalue $1$. The requirement for $P^-=(Q^-_1)^2$
to be invariant under $S$ imposes some restrictions on the permutations.
In fact, we will require that $Q^-_1$ be invariant under $S$, in order
to guarantee that $P^-$ is invariant.

The form of the supercharge from \cite{alpp98} is
\begin{equation}
Q^-_{\alpha} =  \int_0^{\infty}
  [...]b^{\dagger}_{\alpha}(k_3)a_{I}(k_1)a_{I}(k_2) +...+
(\beta_I \beta_J^T - \beta_J \beta_I^T )_{\alpha \beta}
  [...]  b^{\dagger}_{\beta}(k_3)a_{I}(k_1)a_{J}(k_2) + ...
\label{Qminus}
\end{equation}
Here the $\beta_I$ are $8\times 8$ real matrices satisfying
$\{\beta_I,\beta_J^T \} = 2\delta_{IJ}$. We use a special representation for
these matrices given in \cite{GSW}.

Let us consider the expression for $Q^-_1$. The first part of the supercharge
(the one which does not include $\beta$ matrices) is invariant under
(\ref{initsymm}) as long as $g(1)=1$ and $Q[1]=1$. We will consider only such
transformations.
In order to analyze the symmetries of the $\beta$ terms, let
us make the following observation. In the representation of $\beta$ matrices
we have chosen, the expression
${\cal B}^\alpha_{IJ}=
\left(\beta_I\beta^T_J-\beta_J\beta^T_I\right)_{1\alpha} $
may take only the values $\pm 2$ or zero. Moreover, for any pair 
$(I,J)$ there is
at most one value of $\alpha$ corresponding to nonzero ${\cal B}$. This fact
allows us to represent ${\cal B}$ in a compact form. To do so, we
introduce a new object $\mu$ defined by
{\small
\begin{equation}
\mu_{IJ}=\left\{\begin{array}{rl} \alpha\,, & {\cal B}^\alpha_{IJ}=2 \\
                                  -\alpha\,, & {\cal B}^\alpha_{IJ}=-2 \\
                                     0\,,    & {\cal B}^\alpha_{IJ}=0
                                                {\mbox{ for all }}\alpha\,.
                 \end{array}\right.
\end{equation}
}
Our choice of $\beta$ matrices then leads to the following expression for
$\mu$:
{\small
\begin{equation}
\mu=\left(
\begin{array}{rrrrrrrr}
0&5&-7&2&-6&3&-4&8\\
-5&0&-3&6&2&-7&8&4\\
7&3&0&-8&-4&-5&6&2\\
-2&-6&8&0&-5&4&3&7\\
6&-2&4&5&0&-8&-7&3\\
-3&7&5&-4&8&0&-2&6\\
4&-8&-6&-3&7&2&0&5\\
-8&-4&-2&-7&-3&-6&-5&0\\
\end{array}
\right).
\end{equation}
}
We are looking for a subset of transformations (\ref{initsymm}) that satisfy
the conditions $g(1)=1$ and $Q[1]=1$ and leave the matrix $\mu$ invariant. The
latter property means that
\begin{equation}
Q[\mu_{P[I]P[J]}]=g(\mu_{IJ})f(I)f(J)\mu_{IJ}\,.
\label{SymmFinalCond}
\end{equation}
Since the subset of transformations that we seek forms a subgroup $R$ of
the permutation group $S_8\times S_8$, it is natural to look for the 
elements of
$R$ that
square to one. In the case of $S_8\times S_8$, it is known that products of
such elements
  generate the whole group, and, as we will show later, the same is 
true for $R$.
One can construct all $Z_2$ symmetries satisfying (\ref{SymmFinalCond}), but
not all of them are independent. In particular if $a$ and $b$ are two such
symmetries then $aba$ is also a $Z_2$ symmetry. By studying different
possibilities we have found that there are $7$ independent $Z_2$ symmetries in
the group $R$, and we have chosen them to be \\
{\small
\begin{center}
\begin{tabular}{|c|c|c|c|c|c|c|c|c|c|c|c|c|c|c|c|c|}
\hline
   &$a_1$&$a_2$&$a_3$&$a_4$&$a_5$&$a_6$&$a_7$&$a_8$&$b_2$&$b_3$&$b_4$&$b_5$&
   $b_6$&$b_7$&$b_8$\\
\hline
1 &$a_7$&$a_3$&$a_2$&$a_6$&$a_8$&$a_4$&$a_1$&$a_5$&$b_2$&$-b_3$&$-b_4$&$-b_6$&
   $-b_5$&$b_8$&$b_7$\\
\hline
2  &$a_3$&$a_6$&$a_1$&$a_5$&$a_4$&$a_2$&$a_8$&$a_7$&$-b_4$&$b_3$&$-b_2$&$-b_5$&
   $b_8$&$-b_7$&$b_6$\\
\hline
3  &$a_8$&$a_7$&$a_6$&$a_5$&$a_4$&$a_3$&$a_2$&$a_1$&$-b_3$&$-b_2$&$b_4$&$-b_5$&
   $b_7$&$b_6$&$-b_8$\\
\hline
4  &$a_5$&$a_4$&$a_8$&$a_2$&$a_1$&$a_7$&$a_6$&$a_3$&$-b_2$&$-b_7$&$b_8$&$b_5$&
   $-b_6$&$-b_3$&$b_4$\\
\hline
5  &$a_8$&$a_3$&$a_2$&$a_7$&$a_6$&$a_5$&$a_4$&$a_1$&$-b_5$&$-b_3$&$b_7$&$-b_2$&
   $b_6$&$b_4$&$-b_8$\\
\hline
6  &$a_5$&$a_8$&$a_7$&$a_6$&$a_1$&$a_4$&$a_3$&$a_2$&$-b_8$&$b_5$&$-b_4$&$b_3$&
   $-b_6$&$b_7$&$-b_2$\\
\hline
7  &$a_4$&$a_6$&$a_8$&$a_1$&$a_7$&$a_2$&$a_5$&$a_3$&$-b_2$&$-b_6$&$b_5$&$b_4$&
   $-b_3$&$-b_7$&$b_8$\\
\hline
\end{tabular}
\end{center}}
Using Mathematica we explicitly constructed all the symmetries of the type
(\ref{initsymm}) satisfying (\ref{SymmFinalCond}). We found that the group 
of such transformations has $168$ elements, and we have shown that all
of them can be generated from the seven $Z_2$ symmetries mentioned above.

In our numerical procedure we use the $Z_2$ symmetries in the following way.
Since all states relevant for the correlator are singlets under the 
symmetry group
$R$, we join our states in classes and treat the whole class as a new state.
For instance, the simplest nontrivial singlet looks like
\begin{equation}
|1\rangle=\frac{1}{8}\sum_{I=1}^8 {\mbox{tr}}
\left(a^\dagger(1,I)a^\dagger(K-1,I)\right)|0\rangle.
\end{equation}
This means that if, during the construction of the basis, we 
encounter the state
$a^\dagger(1,1)a^\dagger(K-1,1)|0\rangle$ it will be replaced by the class
representative (in this case, by the state $|1\rangle$). Such a procedure
significantly decreases the size of the basis, while keeping all the 
information
necessary for calculating the correlator.

\section{Numerical Results}

Our numerical results are presented in Figs.~1(a) and 1(b).
Figure~1(a) is a log-log plot of $r^4$ times the correlator versus $r$,
so that a $1/r^4$ behavior appears as a flat line and a $1/r^5$
behavior gives rise to a
line with slope $-1$. In Fig.~1(b) we plot the log-log derivative,
which is computed from explicit differentiation inside the sum
and amounts to a replacement of $K_4(M_ir)$ by $M_iK_3(M_ir)$.

Computing this correlator beyond the small-$r$ asymptotics
represents a formidable technical challenge. In
\cite{alpp98} we were able to construct the mass matrix explicitly and
compute the spectrum for
$K=2$, $K=3$, and $K=4$. Even for these modest values of the harmonic
resolution, the Hilbert space contained thousands of states. Previously
in \cite{ahlp99} we used this spectrum and the associated wave function
to calculate the correlator beyond the small-$r$ region. In the
calculation we present here we have made three improvements which have
allowed us to expand the space by a factor of approximately 1000. The first and
most straightforward improvement was to rewrite the code in C++, which
simply runs faster than the Mathematica code and can be exported to faster
machines. The second was to use the discrete flavor symmetry to reduce
the size of the problem at a given resolution.
The third improvement is a numerical algorithm that replaces the
explicit diagonalization with an efficient but accurate approximation.

This numerical algorithm follows from the observation that the
contributions to the eigenstate sum are weighted by the square of the
projection $\langle i|T^{++}(-K)|0\rangle$.  The Lanczos diagonalization
algorithm \cite{Lanczos} will naturally generate the states with
nonzero projection if $T^{++}(-K)|0\rangle$ is used as the starting
vector.  Let $|u_1\rangle$ be the normalized vector proportional
to $T^{++}(-K)|0\rangle$, set $b_1=0$, and construct a sequence
of normalized vectors $|u_n\rangle$ according to the Lanczos
iteration
$b_{n+1}|u_{n+1}\rangle=P^-|u_n\rangle-a_n|u_n\rangle-b_n|u_{n-1}\rangle$,
with $a_n=\langle u_n|P^-|u_n\rangle$.  The $|u_n\rangle$ form an
orthonormal basis with respect to which $P^-$ is tridiagonal and easily
exponentiated.  Because all of these vectors are generated by
applying powers of $P^-$ to $|u_1\rangle$, only those eigenvectors
with nonzero projections on $|u_1\rangle$ can appear.
Although generating a complete basis by iteration can yield
the exact answer,\footnote{Both this statement about the complete
basis and the previous statement about nonzero projections will
hold only in exact arithmetic.  Round-off errors will eventually
destroy these relationships as the Lanczos iteration
proceeds.} doing many fewer iterations, even 20,
can be sufficient to capture the important contributions.
Such an approach to the computation of a matrix element is related
to work by Haydock \cite{Haydock} and others \cite{Others}
on matrix elements of resolvents.

\begin{figure}
\begin{tabular}{cc}
\psfig{file=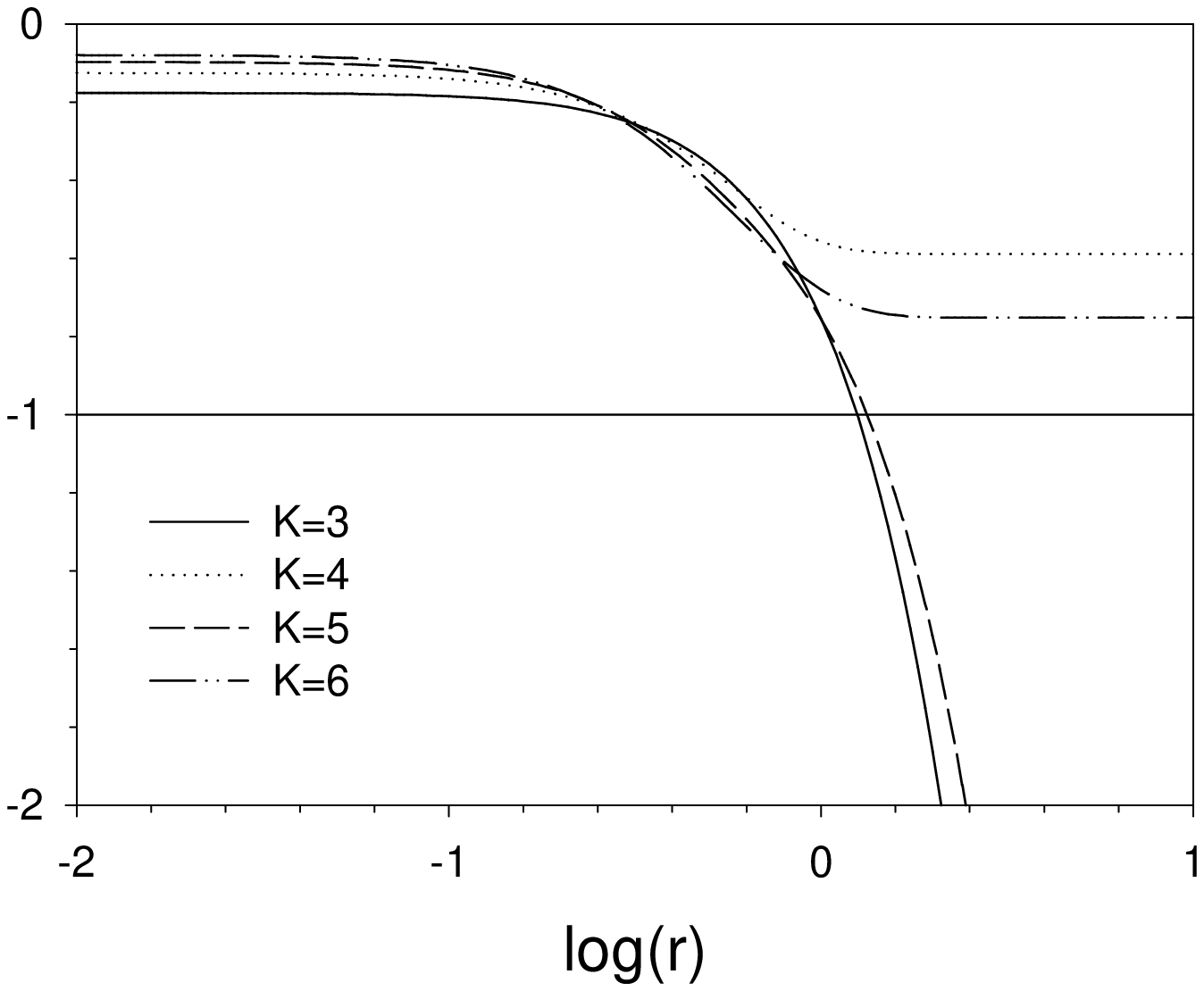,width=3in}  &
\psfig{file=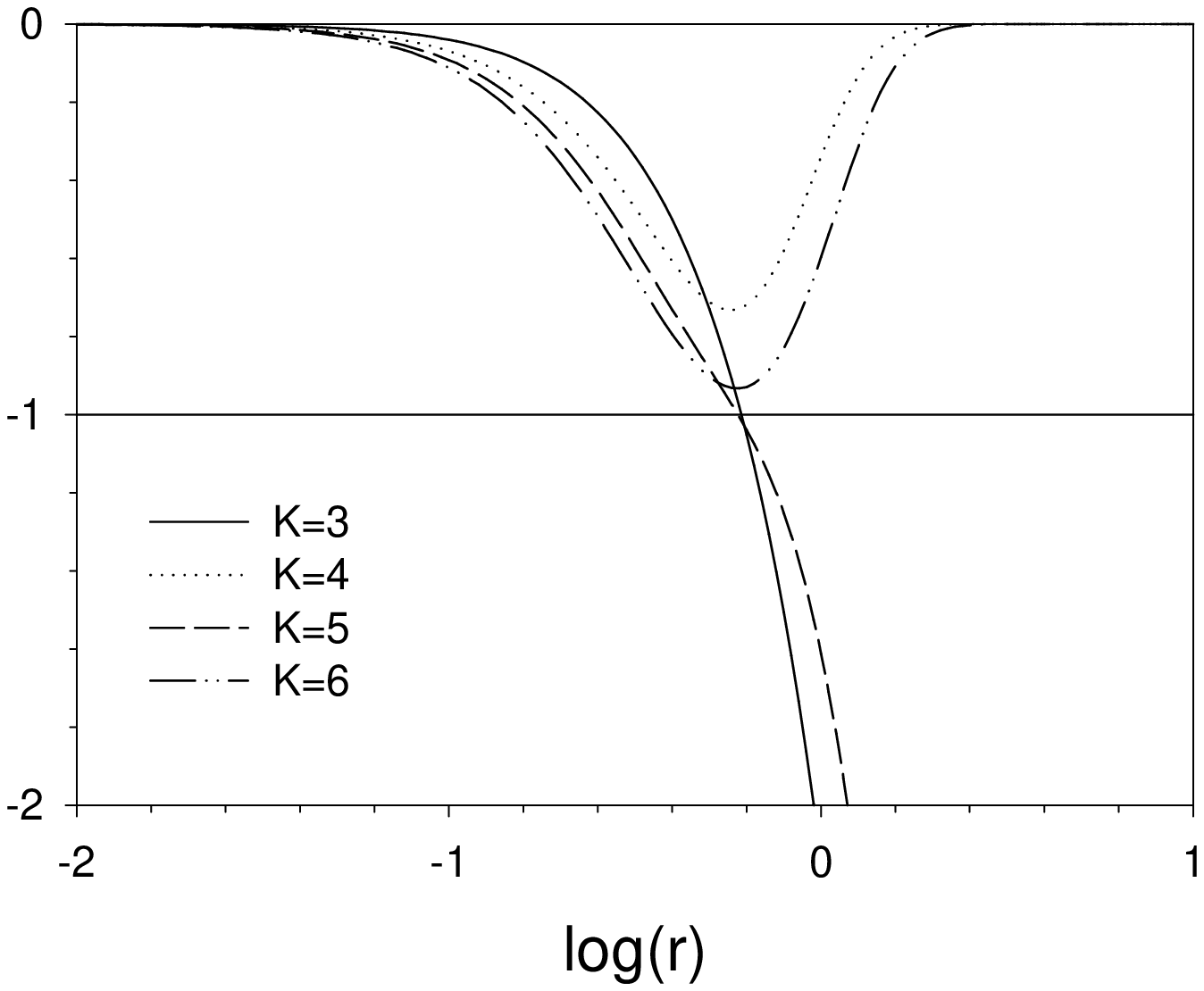,width=3in}  \\
(a) & (b)
\end{tabular}
\caption{(a)the Log-Log
plot of the correlation function $\langle T^{++}(x) T^{++}(0) \rangle
\left({x^- \over x^+} \right)^2 {4 \pi^2 r^4 \over N_c^2 (2 n_b +n_f)}$
v.s. $r$ in the units where $g_{YM}^2 N_c /\pi = 1$ for $K=3,4, 5$ and
$6$ (b) the log-log
derivative with respect to r of the correlation function in (a).
\label{fig}}
\end{figure}

Before discussing our results we need to address the question of massless
states. Our SDLCQ calculation of the spectrum of the (8,8) theory saw
massless states
\cite{alpp98}, and we argued that they were not normalizable bound
states. The argument in that paper was not completely correct but the
conclusion remains true. We find that in these massless states the
number of partons in all the contributions is either all even or all odd
depending on whether the resolution is even or odd.

We have not, however, removed these unphysical states from the data
sets but rather used them to obtain an estimate of the region in $r$ where
the calculation breaks down.
This region is where the unphysical massless states dominate the
correlator sum.  Unfortunately, this is also the region where we
expect the true large-$r$ behavior to dominate the correlator, if
only the extra states were absent.
The correlator is only sensitive to the two-particle
content of the wave function, and we see in Fig.~1(b) the
characteristic behavior  of the massless states at large $r$ only at even
resolutions. In Fig.~1(a) for even resolution, the region where the
correlator  starts to behave like
$1/r^4$ is clearly visible. In Fig.~1(b) we see that for even
resolution the effect of the massless state on the derivative
is felt at smaller values
of $r$ where the even resolution curves start to turn up. We use these
smaller values to estimate the value of $r$ where the large-$N_c$
approximation breaks down. We see that the value increases as we
increase the resolution, as expected. Another estimate of where this
approximation breaks down, that gives consistent values, is the set of
points where the even and odd resolution derivative curves cross. We do
not expect these curves to cross on general grounds, based on work
in \cite{ahlp99}, where we considered a number of other theories.

A proof of the Maldacena conjecture would show up in Fig.~1(b) as a
set of derivative curves that approached and then touched the line
at $-1$ as
we increased the resolution. Convergence in the resolution, $K$, would
appear as a flattening of the derivative curves at $-1$ for the highest
values of $K$.

We see that the derivative curves are approaching $-1$ as we increase the
resolution and appear to be within $10-15 \%$ before the approximation
breaks down. There is however no indication of convergence yet; therefore,
we cannot claim a numerical proof of the Maldacena conjecture.

\section{Conclusion}

In this article, we used the SDLCQ prescription for computing the
correlation function of the stress-energy tensor $T^{++}$, which may be
readily compared with predictions provided by a supergravity analysis
following the conjecture of Maldacena \cite{adscft}. Such a comparison requires
non-perturbative methods on the field theory side, and the SDLCQ approach
is the only numerical method suited to this task. At the present time
the calculation gives results that are within $10-15\%$ of the predicted
value; however, higher resolution calculations are needed
to prove convergence. The results we present here
increase the number of states by a factor of 1000 relative to
\cite{ahlp99}. There are currently available methods that we believe could
give us another factor of 100-1000; however, we have noted in our analysis
of our numerical results that most of the contributions to the
matrix element come from a very small number of eigenfunctions. An analytical
understanding of this phenomenon could greatly accelerate the
calculation.

Finally, we note that, in principle, we could study the proper $1/r$
behavior at large $r$ by computing the $1/N_c$ corrections.  In the
past we have computed such corrections in some theories.  However,
in the present case such a computation seems to be a very large
project indeed.

\section*{Acknowledgments}
The authors would like to acknowledge A. Hashimoto for several very useful
conversations. The calculations of matrices were done with a C++ code
written in part by F. Antonuccio. This work was supported in part by
the US Department of Energy.

\end{document}